\documentclass[onecolumn,groupedaddress,showkeys,showpacs]{revtex4-1}

\pdfoutput=1

\usepackage{graphicx}
\usepackage{caption}
\usepackage{subcaption}
\usepackage{undertilde}
\usepackage{url}
\usepackage{comment}

\usepackage{textpos}
\usepackage{tikz}

\usepackage{amssymb}


\begin{document}

\title{FIT-MART: Quantum Magnetism with a Gentle Learning Curve}
\author{Larry Engelhardt}
\email{lengelhardt@fmarion.edu}
\homepage{http://swampfox.fmarion.edu/engelhardt}
\author{Scott C. Garland}
\author{Cameron Rainey}
\author{Ray A. Freeman}
\affiliation{Department of Physics and Astronomy, Francis Marion University, Florence, South Carolina 29501, USA}

\begin{abstract}
We present a new open-source software package, FIT-MART, that allows non-experts to quickly get started simulating quantum magnetism. FIT-MART can be downloaded as a platform-idependent executable Java (JAR) file. It allows the user to define (Heisenberg) Hamiltonians by electronically drawing pictures that represent quantum spins and operators.  Sliders are automatically generated to control the values of the parameters in the model, and when the values change, several plots are updated in real time to display both the resulting energy spectra and the equilibruim magnetic properties. Several experimental data sets for real magnetic molecules are included in FIT-MART to allow easy comparison between simulated and experimental data, and FIT-MART users can also import their own data for analysis and compare the goodness of fit for different models.
\end{abstract}

\maketitle

\section{Motivation}\label{motivation}

For thousands of years, people have known that certain materials are magnetic, but the origin of this phenomenon remained a mystery until the last century.  This is because any description of material magnetism relies on concepts from the (modern) fields of quantum and statistical physics.  Today, this phenomenon is well-understood by condensed matter physicists; but it still remains a mystery to most people, including most undergraduate physics students. This is the problem that we seek to address. To make quantum magnetism accessible to a broader audience, we have created a user-friendly, open-source software package for modeling simple models of quantum magnetism.  This software can be downloaded from the Open Source Physics collection\cite{fitmart} as a platform-independent executable Java (.JAR) file. To get started, a user simply downloads this file and double clicks.

The software introduced here will be referred to with the acronym FIT-MART which stands for ``Fully-Integrated Tool for Magnetic Analysis in Research and Teaching''. It is a ``fully integrated tool'' in the sense that the entire modeling process is self-contained within one program.  There is no need to first compile the code; results are immediately provided in the form of several plots; and these data can be directly compared to experimental data---all within a single window on a computer. (Results can also easily be exported as images or text files for later use.) This software is also sufficiently sophisticated that it can provide non-trivial results for certain types of real magnetic molecules, and it is currently used by several researchers who study molecular magnetism. To complete the acronym, FIT-MART has been designed to be a viable instructional tool for undergraduate courses (e.g., modern physics, statistical mechanics, and condensed matter physics), even for students who do not have previous computational experience. This is demonstrated for a few specific examples in Sec.~\ref{modeling}.

It is important to note that many advanced simulations are available for researchers who study condensed matter systems. In the field of magnetism, the ALPS project\cite{alps_url, alps1, alps2} provides very powerful codes and libraries for simulating a wide variety of quantum lattice models (Heisenberg, Hubbard, etc.)~using several different numerical methods (exact diagonalization, quantum Monte Carlo, Density Matrix Renormalization Group, etc.). This versatility of models and methods gives ALPS a major advantage compared to FIT-MART, but this versatility comes at a price: FIT-MART is much simpler to use. The strengths and limitations of FIT-MART are explored below in Sec.~\ref{implementation}.

\section{Implementation}\label{implementation}

In order to achieve simplicity in FIT-MART---yet still provide non-trivial results---all simulations use models that are described by the Heisenberg Hamiltonian, 
\begin{equation}\label{heis}
\utilde{\mathcal{H}} = \sum_{\langle i, j \rangle} J_{i,j} \utilde{\vec{s}_i} \cdot \utilde{\vec{s}_j} - g\mu_B\vec{H}\cdot \sum_i \utilde{\vec{s}_i},
\end{equation}
where tildes denote quantum operators. Here $\utilde{\vec{s}_i}$ and $\utilde{\vec{s}_j}$ represent the dimensionless spin operators\footnote{The spin operators in Eq.~(\ref{heis}) have been divided by $\hbar$ such that $J_{i,j}$ has dimensions of energy.} corresponding to two sites, $i$ and $j$, and $J_{i,j}$ describes the strength of the interaction (bond) between these two sites ($J>0$ for an antiferromagnetic interaction).  The second term gives rise to the Zeeman effect: $g$ is the spectroscopic splitting factor, $\mu_B$ is the Bohr-magneton, and $\vec{H}$ is the external magnetic field. Defining $\vec{H}$ to point along the z-axis, the second term becomes $-g\mu_B H\utilde{S}_z$, where $\utilde{S}_z = \sum_i{\utilde{s_{iz}}}$ and $\utilde{s_{iz}}$ represents the z component of the individual spin operator for site $i$. This Hamiltonian commutes with both of the total spin operators, $\utilde{\vec{S}}^2$ 
(where $\utilde{\vec{S}} = \sum_i{\utilde{\vec{s}_i}}$)
and $\utilde{S}_z$, which have the familiar eigenvalues $S(S+1)$ and $M_S$, respectively.
To produce results, we generate the matrix representation of Eq.~(\ref{heis}) using the basis wherein the individual spin operators, $\utilde{{s}_{iz}}$, are diagonal, and then exact diagonalization is used to calculate both the energy eigenvalues and eigenvectors. Since $\utilde{S}_z$ commutues with $\utilde{\mathcal{H}}$, we are able to separately treat each subspace of $\utilde{\mathcal{H}}$ that has a different $M_S$ eigenvalue. This reduces the size of the matrices to be diagonalized, but FIT-MART is still limited to relatively small numbers of spins.

To demonstrate the practical limations of FIT-MART for current personal computers, we have measured computation times, $t$, for several different system sizes.\footnote{Times represent the mean (standard deviation of the mean) delay between changing parameter values (input) and  obtaining updated plots (output). These computations were each timed (repeatedly) using a MacBook Air laptop with a 1.7 GHz Intel Core i5 processor, with 2 GB of RAM allocated to the Java Virtual Machine.} These times are shown in Table \ref{table}, where $N$ represents the number of spins, each of spin $s$. The dimension of the total Hilbert space is then $D=(2s+1)^N$, and $D^\prime$ is the dimension of the largest subspace---i.e., the largest matrix to be diagonalized is $D^\prime \times D^\prime$. When $t\ll 1$ sec, there is no perceptible time delay between the user dragging a slider (e.g., to update $J_{i,j}$) and the plots updating. Hence, FIT-MART runs smoothly for $t\ll 1$ sec,
and there is a noticeable delay for larger system sizes. For $D^{\prime} \gtrsim 1000$, the calculation of all of the energy eigenvalues requires more than 2 GB of RAM, so these computations become impossible on computers that do not have sufficient memory available.
To see exactly how FIT-MART is implemented, we encourage the interested reader to download the source code from Ref.~\cite{fitmart}. FIT-MART was created using Easy Java Simulations  which is a Java code generator that is popular in computational physics education.\cite{ejsSPORE, ejsTPT, ejsTutorial}

\begin{table*}[h]
\caption{Computation times for different system sizes.[6]}
\centering
\begin{tabular}{|c|c|c|c|l|@{\hskip 1.0cm}|c|c|c|c|l|}
	\hline
$s$ &  $N$    &   $D=(2s+1)^N$  & $D^\prime$   & Time (sec)  &
$s$ &  $N$    &   $D=(2s+1)^N$  & $D^\prime$   & Time (sec)\\
	\hline
$1/2$	& 9 		& 512 	& 126  	& $0.256(2)$	& 
3/2 		& 5 		& 1024 	& 155	& $0.526(6)$	\\
$1/2$	& 10 	& 1024 	& 252 	& $0.80(5)$		&
3/2 		& 6 		& 4096 	& 580	& $8.9(2)$		\\
$1/2$	& 11 	& 2048 	& 462  	& $3.8(5)$&
2 		& 3 		& 125 	& 19		& $0.078(5)$ \\
1   		& 5 		& 243 	& 51  	& $0.15(1)$ &
2 		& 4 		& 625 	& 85		& $0.32(1)$ \\
1   		& 6 		& 729 	& 141 	& $0.362(3)$		&
2 		& 5 		& 3125 	& 381	& $2.38(3)$		\\
1   		& 7 		& 2187 	& 393	& $2.06(3)$	&
5/2		& 4		& 1296	& 146	& $0.71(2)$ \\
3/2 		& 4 		& 256 	& 44		& $0.17(2)$&
5/2		& 5		& 7776	& 780	& $31(2)$ \\
\hline
\end{tabular}\label{table}
\end{table*}

\section{Modeling}\label{modeling}

\begin{figure*}[h]
\begin{center}
\begin{tabular}{cc}
& \\
{\includegraphics[width=1.5in]{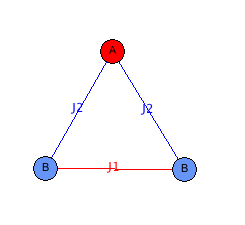}} &
{\includegraphics[width=2.8in]{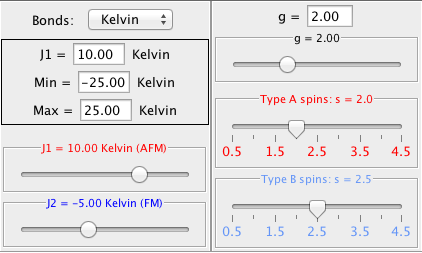}} \\
(a) & (c) \\
& \\
\raisebox{0.5in}{\includegraphics[width=3in]{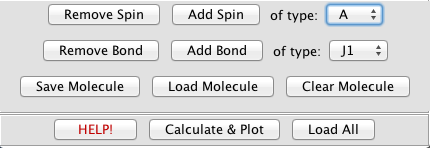}} &
{\includegraphics[width=2.8in]{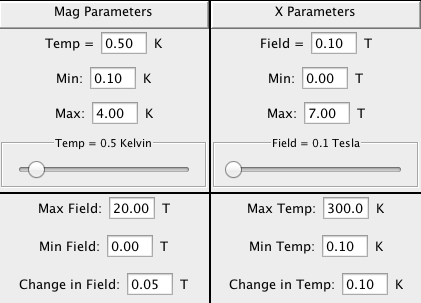}}\\
(b) & (d) \\
\end{tabular}
\caption{
Screenshots from FIT-MART showing how a user defines a model to be simulated.}\label{basic}
\end{center}
\end{figure*}

We now demonstrate how FIT-MART is actually used. Four screenshots from FIT-MART are displayed in Fig.~\ref{basic}. Fig.~\ref{basic}(a) shows a user-drawn picture that corresponds to the Hamiltonian 
\begin{equation}\label{heis2}
\utilde{\mathcal{H}} = J_1 \utilde{\vec{s}_1} \cdot \utilde{\vec{s}_2} + J_2 (\utilde{\vec{s}_1} \cdot \utilde{\vec{s}_3} + \utilde{\vec{s}_2} \cdot \utilde{\vec{s}_3})  - g\mu_B \vec{H}\cdot (\utilde{\vec{s}_{1z}}+\utilde{\vec{s}_{2z}}+\utilde{\vec{s}_{3z}}),
\end{equation}
where the top (red) circle represents $s_3$ and the bottom (blue) circles represent $s_1$ and $s_2$. A user can draw a Hamiltonian such as this by first clicking the buttons shown in Fig.~\ref{basic}(b)---e.g., ``Add Spin'' and ``Add Bond''---and then using a mouse to properly position the spins and bonds within Fig.~\ref{basic}(a). After defining the structure of the Hamiltonian, clicking ``Calculate and Plot'' will produce the controls that are shown in Fig.~\ref{basic}(c), which define the values of the parameters that appear in Eq.~\ref{heis2}; and it will also produce the controls shown in Fig.~\ref{basic}(d), which define the temperatures and fields that are used to calculate equilibrium magnetic properties.

\begin{figure}[t]
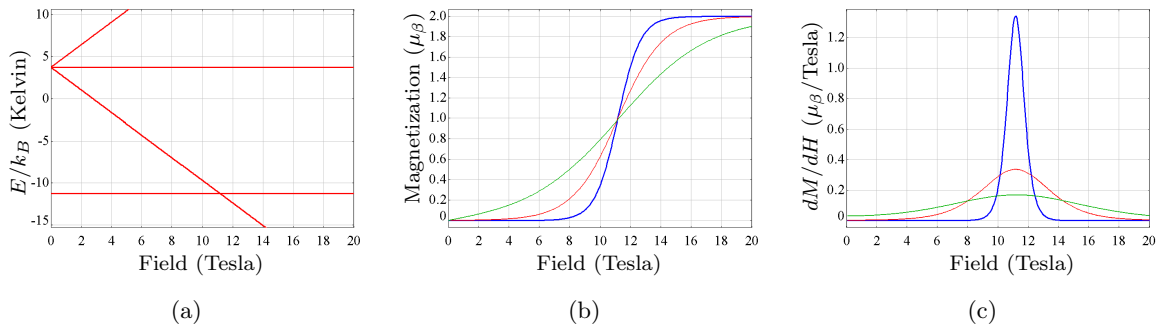

\begin{center}
\begin{tabular}{ccc}
\input{EvsH} &
\input{MvsH} &
\input{dMdHvsH} \\
(a) & (b) & (c) \\
\end{tabular}
\caption{
FIT-MART output for a dimer of spins $s=1/2$ with $J/k_B = 15$ K, and $g=2$. For (b) and (c), curves are shown for three temperatures: 0.5 K (blue), 2 K (red), and 4 K (green).} \label{dimer}
\end{center}
\end{figure}

Fig.~\ref{dimer} shows FIT-MART output for an extremely simple example: two spins, each with $s=1/2$, coupled by an antiferromagnetic bond ($J>0$). This example is pedagogically useful for understanding several topics from introductory quantum and statistical physics. In Fig.~\ref{dimer}(a), we can see an $S=0$ singlet (the zero-field ground state) and an $S=1$ triplet, where the zero-field energy gap\footnote{The user can select between three choices of energy units: meV (=$10^{-3}$ eV), Kelvin, and cm$^{-1}$. These units are discussed in detail in the documentation that is contained within FIT-MART.} between the singlet and the triplet is proportional to $J$. The zero-field degeneracy of the $S=1$ triplet is lifted by the external field with a slope (of $E$ vs.~$H$) that is proportional to $g$, and when $H\approx 11$ Tesla, the ground state changes between $|S=0, M_S=0\rangle$ and $|S=1, M_S=1\rangle$. The effect of this ground state ``level crossing'' is clearly evident in the plots of magnetization, $M$, and differential susceptibility, $dM/dH$, versus field, $H$. There is an abrupt step in $M(H)$ at low temperatures, $T$, and a corresponding narrow peak in $dM/dH$; with increasing $T$, this step (and corresponding peak) are broadened. These features are entirely determined by the Maxwell-Boltzmann distribution function: At very low temperatures, only the ground state is thermally populated; and at elevated temperatures, excited states become increasingly populated. Hence, this example provides a simple platform for student exploration in statistical physics.

In addition to the data that are plotted versus $H$ in Fig.~\ref{dimer}, FIT-MART also calculates and plots three quantities versus temperature for fixed (small) values of $H$: $\chi=M/H$, $T\chi$, and $1/\chi$ (not shown here). For this particular model (two spins $s=1/2$ with antiferromagnetic coupling), the plot of $\chi(T)$ has a peak at $T \approx \frac{2}{3} J/k_B$. A peak in $\chi(T)$ is a typical signature of antiferromagnetism, so it is desirable for students to understand the origin of this peak; and this is easily facilitated with FIT-MART. From Fig.~\ref{dimer}(b), the slope of $M(H)$---which is $\chi$---has a value of $\chi \approx 0$ for $H<5$ Tesla and $T < 2$ Kelvin, and $\chi$ increases with increasing $T$. If you continue to increase $T$ [using the slider labeled ``Temp'' in Fig.~\ref{basic}(d)], you will see that the slope of $M(H)$ continues to increase until $T\approx 9.3$ K and then decreases as $T$ is increased further. These results define the plot of $\chi(T)$, which has a peak at $T \approx 9.3$ K. The plot of $T\chi$ vs.~$T$ increases monotonically with increasing $T$, approaching the Curie constant, $C$, for two independent spin $1/2$ particles as $T\rightarrow \infty$.  In fact, $T\chi > 0.99C$ for $T=300$ K.

\begin{figure}[t]
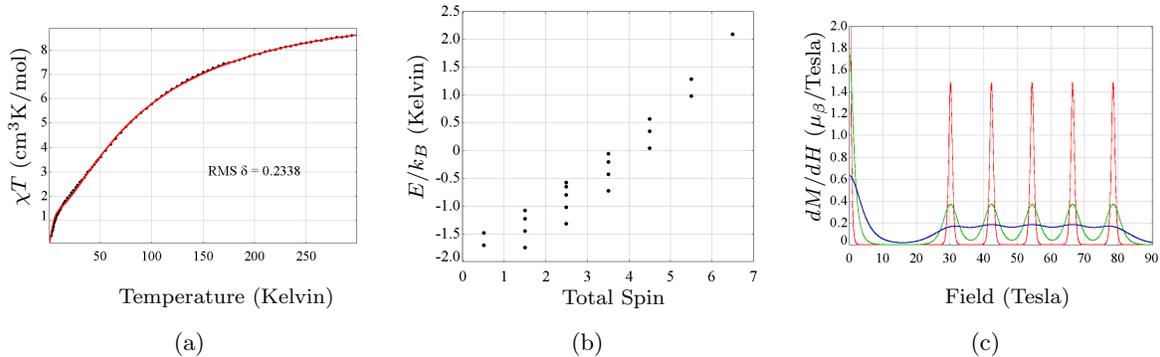

\begin{center}
\begin{tabular}{ccc}
& & \\
\input{xTvT} &
\input{EvS} &
\input{Mn3dMdHvH} \\
(a) & (b) & (c) \\
\end{tabular}
\caption{
FIT-MART output for the Hamiltonian defined by Fig.~\ref{basic}(a) and Eq.~(\ref{heis2}) for $J_1/k_B=9.58$ K, $J_2/k_B=17.04$ K, and $g=2.1$.
(a) Measured\cite{mn3} (symbols) and calculated (curve) $T\chi (T)$ data for a fixed field, $H=7$ Tesla. A diamagnetic correction was included$^4$ as an additional fitting parameter, which yielded the value $\chi_d =1.5\times10^{-3}$ cm$^3$/mol. (b) The zero-field energy spectrum versus the total spin quantum number, $S$, for the model that provides the best fit for $T\chi$.
(c) $dM/dH$ versus $H$ for three temperatures: 0.5 K (red), 2.0 K (green), and 5.0 K (blue).}\label{mn3}
\end{center}
\end{figure}

Fig.~\ref{mn3} shows FIT-MART output for a less simple, experimentally relevant example. Fig.~\ref{mn3}(a) shows published\cite{mn3} $T\chi$ data (symbols) for a \{Mn$_3$\} molecule whose Hamiltonian is defined by Fig.~\ref{basic}(a) and Eq.~(\ref{heis2}). Using the notation of Eq.~(\ref{heis2}), $s_1 = s_2 = 2$ (there are two Mn$^{3+}$ ions), and $s_3=5/2$ (there is one Mn$^{2+}$ ion). The best fit for these data [the solid curve in Fig.~\ref{mn3}(a)] is achieved using the parameters $J_1/k_B=9.58$ K and $J_2/k_B=17.04$ K, and a quantitative measure of the goodness of fit (the root-mean-square deviation between experiment and theory) for this set of parameters is displayed within Fig.~\ref{mn3}(a).  FIT-MART also allows one to include corrections for the bulk materal including diamagnetism and impuritities,\footnote{Including corrections for the bulk material, the magnetic susceptibility in FIT-MART is given by $\chi(T) = \chi_m(T) - \chi_d + M_{imp}(T,H)/H$, where $\chi_m(T)$ is the susceptibility of the molecule and $\chi_d$ is the diamagnetic correction. $M_{imp}(T,H)$ describes the magnetization of $N$ detached (``impurity'') spins, $s$, per molecule, which is defined in terms of the Brillouin function, $B_s(x)$, according to $M_{imp}=Ng\mu_B s B_s\left(\frac{g\mu_BsH}{k_BT}\right)$.} and diamagnetism has been included as a fitting parameter to obtain the data in Fig.~\ref{mn3}(a). This set of parameters produces the zero-field energy spectrum shown in Fig.~\ref{mn3}(b), which has an $S=3/2$ ground state. Fig.~\ref{mn3}(c) contains predictions for high-field, low-temperature magnetization measurements.  If this model is correct, five ground state level crossings should be observable between $H=30$ and 80 Tesla, resulting in the eventual saturation of the magnetization (to an $S=13/2$ ground state) when $H \approx 78$ Tesla. We note that these are new predictions that could be tested with pulsed-field measurements.

\newpage

\section{Summary and future development}\label{motivation}

We have introduced an easy-to-use software package (``FIT-MART'') that allows users to perform advanced quantum mechanics calculations and analyze magnetic data for certain classes of real materials. This software is sufficiently user-friendly that it can be used in undergraduate instruction without previous computational experience. It is also sufficiently powerful that it has been used by several researchers studying magnetic molecules. An important feature of FIT-MART is the inclusion of experimental data, and the real-time analysis of the goodness of fit. [See Fig.~\ref{mn3}(a).] The process of varying model parameters and comparing simulated and experimental data is an important part of scientific discovery, and FIT-MART provides a means for undergraduate physics students to be able to experience this modeling process first hand.\footnote{For a description of an analogous software package that allows users to compare experiment and theory---but in the context of cosmology---see Ref.~\cite{cosmoEJS}.}

In subsequent versions of FIT-MART, the calculation of energy eigenvalues will be parallelized, which will significantly reduce the computation times listed in Table~\ref{table} when run on a multi-core computer. Automated fitting routines will also be incorporated in order to automatically search multidimensional parameter spaces.  Finally, a series of exercises will be included within the FIT-MART documentation in order to support instructors wishing to use FIT-MART within their courses.

\bibliographystyle{elsarticle-num}
\bibliography{csp.bib}

\begin{thebibliography}{10}
\expandafter\ifx\csname url\endcsname\relax
  \def\url#1{\texttt{#1}}\fi
\expandafter\ifx\csname urlprefix\endcsname\relax\def\urlprefix{URL }\fi
\expandafter\ifx\csname href\endcsname\relax
  \def\href#1#2{#2} \def\path#1{#1}\fi

\bibitem{fitmart}
L.~Engelhardt, C.~Garland, C.~Rainey, A.~Freeman, {FIT-MART project},
  \url{http://www.compadre.org/osp/items/detail.cfm?ID=12308}.

\bibitem{alps_url}
M.~Troyer, et~al., {ALPS project}, \url{http://alps.comp-phys.org}.

\bibitem{alps1}
B.~Bauer, et~al., The alps project release 2.0: open source software for
  strongly correlated systems, J.~Stat.~Mech. (2011) P05001.

\bibitem{alps2}
A.~F. Albuquerque, et~al., The alps project release 1.3: open source software
  for strongly correlated systems, J.~Magn.~Magn.~Mater. 310 (2007) 1187.

\bibitem{Note1}
The spin operators in Eq.~(\ref {heis}) have been divided by $\hbar $ such that
  $J_{i,j}$ has dimensions of energy.

\bibitem{Note2}
Times represent the mean (standard deviation of the mean) delay between
  changing parameter values (input) and obtaining updated plots (output). These
  computations were each timed (repeatedly) using a MacBook Air laptop with a
  1.7 GHz Intel Core i5 processor, with 2 GB of RAM allocated to the Java
  Virtual Machine.

\bibitem{ejsSPORE}
W.~Christian, F.~Esquembre, L.~Barbato, Spore award: Open source physics,
  Science 334 (2011) 1077--1078.

\bibitem{ejsTPT}
W.~Christian, F.~Esquembre, Modeling physics with easy java simulations,
  Phys.~Teach. 45 (2007) 475--480.

\bibitem{ejsTutorial}
L.~Engelhardt, {EJS Video Tutorial},
  \url{http://www.compadre.org/OSP/tutorials/EJS_Tutorial}.

\bibitem{Note3}
The user can select between three choices of energy units: meV (=$10^{-3}$ eV),
  Kelvin, and cm$^{-1}$. These units are discussed in detail in the
  documentation that is contained within FIT-MART.

\bibitem{mn3}
B.~J. Suh, D.~Procissi, J.~K. Jung, S.~Bud'ko, Y.~J. Jeon, W.~S.~Kim, D.~Y.
  Jung, Magnetic properties and spin dynamics in magnetic molecule {Mn3},
  J.~Appl.~Phys. 93 (2003) 798--7100.

\bibitem{Note4}
Including corrections for the bulk material, the magnetic susceptibility in
  FIT-MART is given by $\chi (T) = \chi _m(T) - \chi _d + M_{imp}(T,H)/H$,
  where $\chi _m(T)$ is the susceptibility of the molecule and $\chi _d$ is the
  diamagnetic correction. $M_{imp}(T,H)$ describes the magnetization of $N$
  detached (``impurity'') spins, $s$, per molecule, which is defined in terms
  of the Brillouin function, $B_s(x)$, according to $M_{imp}=Ng\mu _B s
  B_s\left ({\begingroup g\mu _BsH\endgroup \over k_BT}\right )$.

\bibitem{Note5}
For a description of an analogous software package that allows users to compare
  experiment and theory---but in the context of cosmology---see Ref.~\cite
  {cosmoEJS}.

\bibitem{cosmoEJS}
J.~Moldenhauer, L.~Engelhardt, K.~Stone, E.~Shuler, Modern cosmology:
  Interactive computer simulations that use recent observational surveys,
  Am.~J.~Phys. (2013, in press);\textrm{ }\url{http://arxiv.org/abs/1212.4661}.

\end{thebibliography}

\end{document}